\documentclass[10pt,a4paper,fleqn]{article}

\usepackage[
    top=20mm,
    bottom=20mm,
    left=18mm,
    right=18mm
]{geometry}

\usepackage{graphicx}
\usepackage{multicol}
\usepackage{titlesec}
\usepackage{fancyhdr}
\usepackage{setspace}
\usepackage{array}
\usepackage{booktabs}
\usepackage{tikz}
\usepackage{tcolorbox}
\usepackage{amssymb}
\usepackage{caption}
\usepackage{fontspec}
\usepackage{amsmath}
\usepackage{graphicx}
\usepackage{siunitx}
\usepackage{natbib}
\usepackage{multirow}

\usepackage{hyperref}
\usepackage{url}
\usepackage{lipsum}

\usepackage{booktabs}
\usepackage{titlesec}
\titleformat{\section}
    {\normalsize\bfseries}
    {\thesection}
    {1em}
    {}

\titleformat{\subsection}
  {\footnotesize\bfseries}
  {\thesubsection}
  {1em}
  {}

\titleformat{\subsubsection}
  {\footnotesize\bfseries}
  {\thesubsubsection}
  {1em}
  {}

\titleformat{\paragraph}[runin]
  {\footnotesize\bfseries}
  {\theparagraph}
  {1em}
  {}

\usepackage{lipsum}

\usepackage{varwidth}
\usepackage{adjustbox}

\definecolor{tether-gray}{HTML}{F2F1EF}

\newtcolorbox[auto counter]{custombox}[2][]{
    title={Box~\thetcbcounter: #2},
    label={#1},
    fonttitle=\bfseries,
    coltitle=black,
    before title=\vspace{15pt},
    colback=tether-gray,
    colframe=tether-gray,
    boxrule=0pt,
    arc=5pt,
    left=15pt,
    right=15pt,
    bottom=15pt,
    width=\textwidth,    
}

\begin{document}


\begin{flushleft}
\includegraphics[height=20pt]{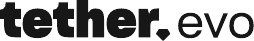}
\end{flushleft}

\vspace{10pt}


\begin{tcolorbox}[
    colback=tether-gray,
    colframe=tether-gray,
    boxrule=0pt,
    arc=5pt,
    left=15pt,
    right=15pt,
    top=15pt,
    bottom=15pt,
    width=\textwidth
]


{\huge\bfseries
Mapping Whisper Representations to Human ECoG Responses with Interpretable Time-Resolved Neural Encoding
}

\vspace{5pt}


{\small\bfseries
Matteo Ciferri\textsuperscript{1},
Tommaso Boccato\textsuperscript{2},
Michal Olak\textsuperscript{2},
Matteo Ferrante\textsuperscript{1,2}\textsuperscript{*},
Nicola Toschi\textsuperscript{1,3}\textsuperscript{*}
}

\vspace{2.5pt}

{\small
\textsuperscript{1}Department of Biomedicine and Prevention, University of Rome Tor Vergata, Rome, Italy\\
\textsuperscript{2}Tether Evo\\
\textsuperscript{3}A.A.
Martinos Center for Biomedical Imaging, Massachusetts General Hospital, Boston, USA\\
\textsuperscript{*} Equal Contribution
}

\vspace{15pt}


{\footnotesize
Understanding how speech foundation models relate to human cortical activity is a key challenge for computational neuroscience. Here, we investigate how internal representations from Whisper predict intracranial ECoG responses during naturalistic speech perception. We introduce a time-resolved neural encoder that combines speech embeddings with a recurrent temporal model and soft attention, allowing us to examine layer-wise brain alignment. Intermediate Whisper layers provide the strongest correspondence with neural activity, supporting a hierarchical match between model representations and cortical speech processing. Comparisons with baselines show that high-resolution ECoG responses benefit from temporally structured modelling beyond linear mappings from the same speech representations. In addition, attention maps reveal temporally local alignment between speech embeddings and neural responses, while a phonemic interpretability analysis identifies anatomically coherent phoneme-category organization among encoding-informative electrodes. Together, these results suggest that speech foundation models offer a useful framework for studying time-resolved cortical speech representations.
}

\end{tcolorbox}

\vspace{10pt}


\footnotesize


\section{Introduction}

Understanding how the human brain encodes the acoustic, phonetic, and linguistic structure of speech is a central challenge in computational neuroscience and cognitive science. Intracranial electrophysiology (ECoG) provides a unique opportunity to study these mechanisms with high temporal and spatial precision, revealing how auditory and language-responsive cortical regions transform continuous acoustic input into increasingly abstract linguistic representations. Recent advances in large-scale language models have shown that their hierarchical internal representations mirror aspects of neural processing, suggesting a promising bridge between machine-learned features and cortical dynamics \citep{raugel2025scaling, antonello2023scaling, oota_deep_2023, ciferri2025reconstructing, ferrante2023eyes, scotti2024mindeye2}.


At the same time, modeling the fine-grained mapping between continuous speech and neural activity remains difficult. Naturalistic listening conditions introduce substantial variability, and neural responses reflect a mixture of low-level acoustic cues, phonetic structure, and higher-order linguistic information unfolding over time. A central methodological question is therefore whether this mapping can be adequately captured by classical linear encoding models, or whether the temporal resolution and signal quality of intracranial recordings justify the use of more expressive, temporally structured models. This question is particularly relevant for ECoG, where high-gamma activity provides a spatially localized and temporally precise measure of cortical responses, potentially exposing nonlinear and time-dependent stimulus-response relationships that may be attenuated or averaged out in lower-resolution modalities. Capturing these dependencies requires approaches that are both temporally aware and capable of leveraging the hierarchical organization of modern speech encoders.

In this work, we investigate how representations from Whisper \citep{radford2023robust}, a large speech recognition model trained at scale, predict ECoG activity during naturalistic speech perception. Using the Podcast ECoG dataset \citep{zada2025podcast}, we construct a time-resolved neural encoding model that aligns Whisper embeddings to cortical responses. This architecture allows us to examine which layers of the speech model best correspond to neural dynamics and how temporal information is integrated during word processing.

Beyond encoding performance, we introduce a phonemic interpretability framework that identifies electrodes with selective responses to articulatory phoneme categories. This analysis reveals anatomically structured clusters across the superior temporal cortex, providing a bridge between model-derived features and known phonetic organization in auditory cortex.

Overall, our results demonstrate that (i) intermediate layers of Whisper best predict neural activity, (ii) the learned temporal attention mechanism captures temporally local stimulus-response structure, and (iii) phoneme-selective responses follow interpretable spatial gradients in the human cortex. Together, these findings highlight how modern speech models provide a powerful lens for understanding the neural basis of language perception.

\section{Related Work}

Early work on the neural basis of speech perception used hypothesis-driven encoding models based on spectro-temporal receptive fields or hand-crafted phonetic features to explain responses in auditory cortex \citep{mesgarani2009influence, chang2010categorical}. 

The advent of self-supervised and large-scale speech models has enabled more powerful, data-driven encoding approaches. Representations from models such as Wav2Vec2 \citep{baevski2020wav2vec} and HuBERT \citep{hsu2021hubert} have been used to predict fMRI, MEG, and ECoG responses, revealing layer-wise correspondences between model hierarchies and auditory pathways, from subcortical nuclei to higher-order language areas \citep{li2023dissecting,anderson2024deep}, such as the superior temporal gyrus (STG). In parallel, invasive speech brain–computer interface (BCI) work has focused primarily on decoding, using high-resolution intracortical or ECoG recordings together with recurrent or transformer architectures to reconstruct text or audio from attempted or imagined speech \citep{willett2023high,metzger2023high,chen2024neural}. These studies demonstrate the feasibility of rich speech decoding, but typically optimize task performance rather than building explicit encoding models of continuous natural language comprehension.

Closer to our setting, several recent works have begun to exploit large speech–language models as joint models of brain and behavior. \citet{goldstein2025unified} show that a unified acoustic-to-speech-to-language embedding space derived from Whisper captures neural activity during everyday conversations across widespread cortical networks. Their analysis, however, focuses on large-scale alignment between model stages and brain regions rather than modeling fine-grained, time-resolved mappings around individual word events.

In contrast, our work explicitly targets word-locked, time-resolved neural encoding, introducing a temporally aware alignment mechanism that enables phoneme level interpretability of the learned speech–brain mappings.

\section{Materials and Methods}

\begin{figure*}
    \centering
    \includegraphics[width=.99\linewidth]{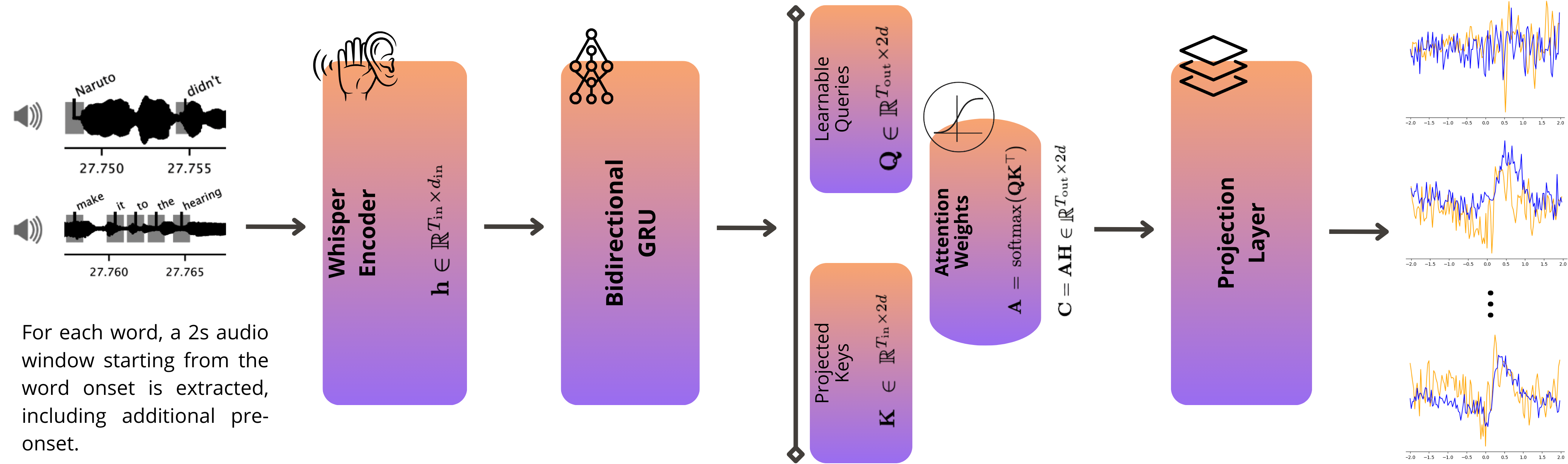}
    \caption{Overview of the neural encoding architecture. Word-aligned speech segments are processed by the Whisper encoder and a bidirectional GRU, followed by a temporal attention mechanism that maps speech representations to neural timepoints and a linear projection that predicts ECoG activity.}
    \label{fig:neural_model_ecog}
\end{figure*}

In the following sections we describe the dataset used and propose the brain encoding framework to estimate neural activity from embeddings of a natural speech stimulus (Figure \ref{fig:neural_model_ecog}). The goal is to extract meaningful auditory information from the most responsive neural channels. 

The dataset is freely accessible on the OpenNeuro platform (dataset number "\textit{ds005574}"). Implementation code for reproducibility is available at the anonymised repository: \url{https://anonymous.4open.science/r/ecog_encoding-9A21/}. All experiments were carried out on a high performance server equipped with eight NVIDIA A100 GPUs (80 GB each, interconnected via NVLINK), 256 CPU threads and 2 TB of system memory.

\subsection{Dataset}
We analyzed the publicly available Podcast ECoG Dataset \citep{zada2025podcast}, which contains intracranial electrophysiological recordings (ECoG) acquired from 9 subjects while listening to a naturalistic 30-minute spoken narrative comprising 5{,}137 words. The dataset includes a total of 1{,}330 electrodes covering auditory, premotor, and higher-order language regions across participants. Neural signals were sampled at 512~Hz and stored in BIDS-compatible structures \citep{gorgolewski2016brain}.  
During listening, participants heard the continuous podcast ``Monkey in the Middle'', and no behavioral task was required. This naturalistic setting enables the study of spontaneous cortical responses to ecologically valid speech.


We used the high-gamma ECoG derivatives, following the preprocessing pipeline released with the Podcast ECoG dataset (including automated artifact handling and high-gamma range 70–200 Hz filtering). Neural data were subsequently downsampled from 512~Hz to 32~Hz using MNE’s FFT-based resampling procedure, which applies an implicit low-pass filter prior to resampling to prevent aliasing, as in the official tutorial \citep{zada2025podcast}.
Finally, neural signals were z-scored channel-wise using normalization statistics computed on the training data and applied to the held-out test data. This preprocessing substantially reduces computational complexity while preserving the temporal resolution necessary to capture word-level cortical dynamics unfolding over hundreds of milliseconds.

Since both neural and audio data are continuous signals, we defined each data sample as a fixed temporal segment centered on a spoken word. Specifically, for each word onset provided in the time-aligned transcript, we extracted a neural segment spanning from -2 s to +2 s relative to onset. The corresponding audio segment was aligned to the same onset but included a shorter pre-onset context of 200 ms, avoiding the inclusion of temporally distant speech segments that are unlikely to contribute to the word-locked neural response, while retaining the post-onset portion corresponding to the same temporal window. The resulting word-aligned neural and audio segments were then paired to form a collection of (audio, neural) samples used for model training and evaluation.

\subsection{Audio Processing}
Audio segments were resampled to 16~kHz and passed to the Whisper-base encoder \citep{radford2023robust} as a frozen feature extractor. Encoder hidden states were extracted at the model’s intrinsic temporal resolution of approximately 50~Hz (one frame every $\sim$20~ms). Whisper is designed to process audio segments of 30~s in duration; in our setting we provide shorter, word-aligned segments. Accordingly, for each input we retained the first $50 \times (T_{\mathrm{pre}} + T_{\mathrm{post}})$ encoder timesteps, corresponding to the temporal extent of the segmented audio window (e.g., 110 timesteps for a 2.2~s segment). 

For each segment $x(t)$, the $n$-th hidden layer of Whisper provided a sequence of latent speech embeddings: $\mathbf{h}_t = f_{\text{Whisper}}^{(n)}(x_t).$
Prior analyses suggest that Whisper representations may capture information at multiple levels of abstraction across encoder layers \citep{goldstein2025unified}: early layers predominantly encode the acoustic envelope and fine-grained spectral structure, whereas deeper layers progressively encode phonetic and articulatory information. Because Whisper retains the temporal structure of the input, its hidden states constitute an expressive basis for modeling time-resolved brain activity.

\subsection{Neural Encoding Architecture}

In order to model the mapping from hierarchical Whisper representations to neural activity, we adopted a time-aware neural encoder composed of a bidirectional Gated Recurrent Unit (GRU) followed by a temporal attention mechanism and a linear projection stage.

Let $\mathbf{h} \in \mathbb{R}^{T_{\mathrm{in}} \times d_{\mathrm{in}}}$ denote the sequence of Whisper embeddings for a speech segment, where
$T_{\mathrm{in}}$ is the number of input timepoints and $d_{\mathrm{in}}$ is the input dimensionality. The embeddings are first
processed by a bidirectional GRU with hidden size $d$, yielding the
contextualized representation $ \mathbf{H} = \text{BiGRU}(\mathbf{h}) \in \mathbb{R}^{T_{\mathrm{in}} \times 2d}, $ where each row $\mathbf{H}_{t}$ corresponds to the concatenated forward and backward hidden states at time $t$. 

To align these contextualized speech embeddings with the neural response
over time, we introduced a learnable soft alignment mechanism. The model
learns a set of $T_{\mathrm{out}}$ temporal queries, $ \mathbf{Q} \in \mathbb{R}^{T_{\mathrm{out}} \times 2d}, $ where $T_{\mathrm{out}}$ denotes the number of neural timepoints to be predicted. The contextualized representations are projected into a key space, $ \mathbf{K} \in \mathbb{R}^{T_{\mathrm{in}} \times 2d}, $ and attention weights are obtained by $ \mathbf{A} = \mathrm{softmax}\!\left( \mathbf{Q}\mathbf{K}^{\top} \right) \in \mathbb{R}^{T_{\mathrm{out}} \times T_{\mathrm{in}}}, $ where the softmax is applied along the input-time dimension $T_{\mathrm{in}}$. Each row of $\mathbf{A}$ distributes attention over the Whisper time axis for a given predicted neural time index.

The resulting context vectors are computed as $ \mathbf{C} = \mathbf{A}\mathbf{H} \in \mathbb{R}^{T_{\mathrm{out}} \times 2d}, $ providing a soft temporal alignment of speech features to neural
timepoints. This mechanism allows the model to integrate temporally
distant acoustic or phonetic cues when predicting the neural signal. Finally, the context vectors are mapped onto predicted neural activity by a
linear projection $ \hat{\mathbf{Y}} = \mathbf{C}\mathbf{W}_{\mathrm{out}}^\top
\in \mathbb{R}^{ T_{\mathrm{out}} \times S},$ where $S$ is the number of electrodes. 

The model was trained using a contrastive objective that aligns predicted and ground-truth neural responses. Predicted and real ECoG responses were compared using a temperature-scaled cosine similarity loss, encouraging each prediction to be most similar to its corresponding target within a batch. The temperature parameter was learned jointly with the model.


We chose a recurrent backbone to explicitly model short-range temporal dependencies while keeping the architecture lightweight. Interpretability is introduced through the temporal attention mechanism, which provides a transparent weighting of speech embeddings when predicting neural responses. More expressive architectures were avoided to reduce overfitting, given the limited number of word-aligned training samples.

\subsection{Evaluation}
Model evaluation was performed using 4-fold cross-validation with temporally contiguous splits. Specifically, the continuous audio stream was partitioned into five non-overlapping segments, and in each fold one segment was held out for testing while the remaining segments were used for training. This strategy was adopted to prevent potential data leakage arising from temporal autocorrelations in naturalistic speech, and to avoid inflated performance that could result from the model interpolating between temporally adjacent and highly similar samples.

Model performance was quantified using Pearson's correlation between real and predicted neural activity: $r = \frac{\mathrm{Cov}(\hat{\mathbf{y}},\mathbf{y})}{\sigma_{\hat{\mathbf{y}}}\,\sigma_{\mathbf{y}}}.$
For each subject and Whisper layer, we computed $r$ for each electrode and neural timepoint across held-out test samples, and then averaged scores across cross-validation folds. Neural timepoints were defined relative to word onset, with negative lags corresponding to pre-onset activity and positive lags corresponding to post-onset responses.

\subsection{Phonemic Interpretability}
In order to better understand how linguistic elements are represented in neural activity, a multi-stage interpretability framework was implemented. This module identifies significant neural activations time-locked to words, relates them to underlying phonemic categories, and organizes the results into spatial clusters on the cortical surface. The clusters are extracted using real neural data with only significant channels from the encoding results.

Neural activation significance was assessed using a sliding-window analysis applied to z-scored ECoG time series aligned with each spoken word. For each word-channel pair ($w,s$), the mean z-score within each window was extracted. A window was considered significant if the summary statistic (the mean) exceeded a fixed threshold $z_{thr}$. For each significant window, the peak index $t^\ast = \arg\max_t |\bar{z}_{w,s}(t)|$ was stored, representing the most responsive neural moment relative to the linguistic event and channel. This approach is conceptually related to recent analyses of phoneme-selective activation patterns in Whisper encoder representations, where a similar strategy is used to characterize category-specific temporal peaks in model layer activations.

Each spoken word was decomposed into phonemic sequences using a grapheme-to-phoneme (G2P) model \citep{bisani2008joint}. Each phoneme was assigned to a broader articulatory category (following the \citet{willett2023high} results) such as front vowels, alveolar consonants, bilabial consonants, diphthongs, and others (Table \ref{tab:phoneme_categories}). 

\begin{table}[t]
\centering
\caption{Phoneme categories used in the interpretability analysis. Phonemes are reported in ARPAbet notation.}
\label{tab:phoneme_categories}
\begin{tabular}{ll}
\toprule
\textbf{Category} & \textbf{Phonemes} \\
\midrule
Vowels (Front)      & IY, IH, EY, EH \\
Vowels (Central)    & AH, ER \\
Vowels (Back)       & UW, UH, AO, AA \\
Diphthongs          & AY, OY, AW \\
\midrule
Bilabial            & P, B, M \\
Alveolar            & T, D, N, S, Z \\
Velar--Palatal      & K, G, NG \\
Post-Alveolar       & SH, ZH \\
Labio-Dental        & F, V \\
Dental / Glottal    & TH, DH, HH \\
Approximants        & L, R, W, Y \\
\bottomrule
\end{tabular}
\end{table}

We quantified category-specific phoneme selectivity at each channel using a multinomial likelihood ratio test (G-test). For each channel, we compared the observed distribution of phoneme categories to the global distribution computed across the entire dataset. Let $O_{sc}$ be the observed count of category $c$ in channel $s$, and $E_{sc} = n_s p_c$ the expected count under the global proportions $p_c$. The goodness-of-fit statistic is:
$G_s = 2 \sum_{c} O_{sc} \log\left(\frac{O_{sc}}{E_{sc}}\right).$
Channels with large $G_s$ deviate significantly from the global phoneme distribution. To assess which categories drive this deviation, we used the per-category contribution $G_{sc}$ and computed one-sided chi-square $p$-values (denoted by $p_{s,c}$) testing for over-representation.

For each channel, we defined the set of significantly over-represented categories as $C_s = \{ c \;|\; p_{s,c} < \alpha,\; O_{sc} > E_{sc} \},$ and assigned an initial label $c_s^{\ast} = \arg\min_{c \in C_s} p_{s,c}.$
To ensure anatomical coherence, we applied a spatial refinement step based on
majority voting among neighbors in MNI space. For each channel $s$, we
identified all electrodes $N_r(s)$ within a fixed Euclidean radius and reassigned its label according to $c_s^{\mathrm{refined}} = \arg\max_{c \in C_s} \left|\{\, j \in N_r(s) \;|\; c_j^{\ast} = c \}\right|,$ while restricting the vote to categories in $C_s$ to ensure that no label lacking local statistical support could be assigned.

In order to quantify this organization, we compared the spatial coherence of phoneme-category clusters obtained from the subset of encoding-selected electrodes (most responsive channels) against clusters obtained by running the same pipeline on all electrodes. We used complementary clustering metrics that capture separation, compactness, and local consistency. The Silhouette score \citep{rousseeuw1987silhouettes} quantifies how well electrodes assigned to the same phoneme category are separated from electrodes assigned to other categories, with higher values indicating better-defined clusters. The Davies--Bouldin index \citep{davies2009cluster} measures the ratio between within-cluster dispersion and between-cluster separation, where lower values reflect more compact and well-separated clusters. Finally, local label entropy (Shannon entropy) quantifies the diversity of phoneme labels within a spatial neighborhood of each electrode, with lower entropy indicating greater local homogeneity and cleaner anatomical organization.

\section{Results}

We first evaluate how well hierarchical speech embeddings predict time-resolved ECoG responses during naturalistic speech perception, revealing a structured correspondence between Whisper layers and neural dynamics.
Finally, we assess the interpretability of the learned mappings, showing that the model captures meaningful temporal alignment and recovers anatomically coherent, phoneme-selective cortical organization.

\subsection{Encoding Performance}

\begin{figure*}
    \centering
    \includegraphics[width=.96\linewidth]{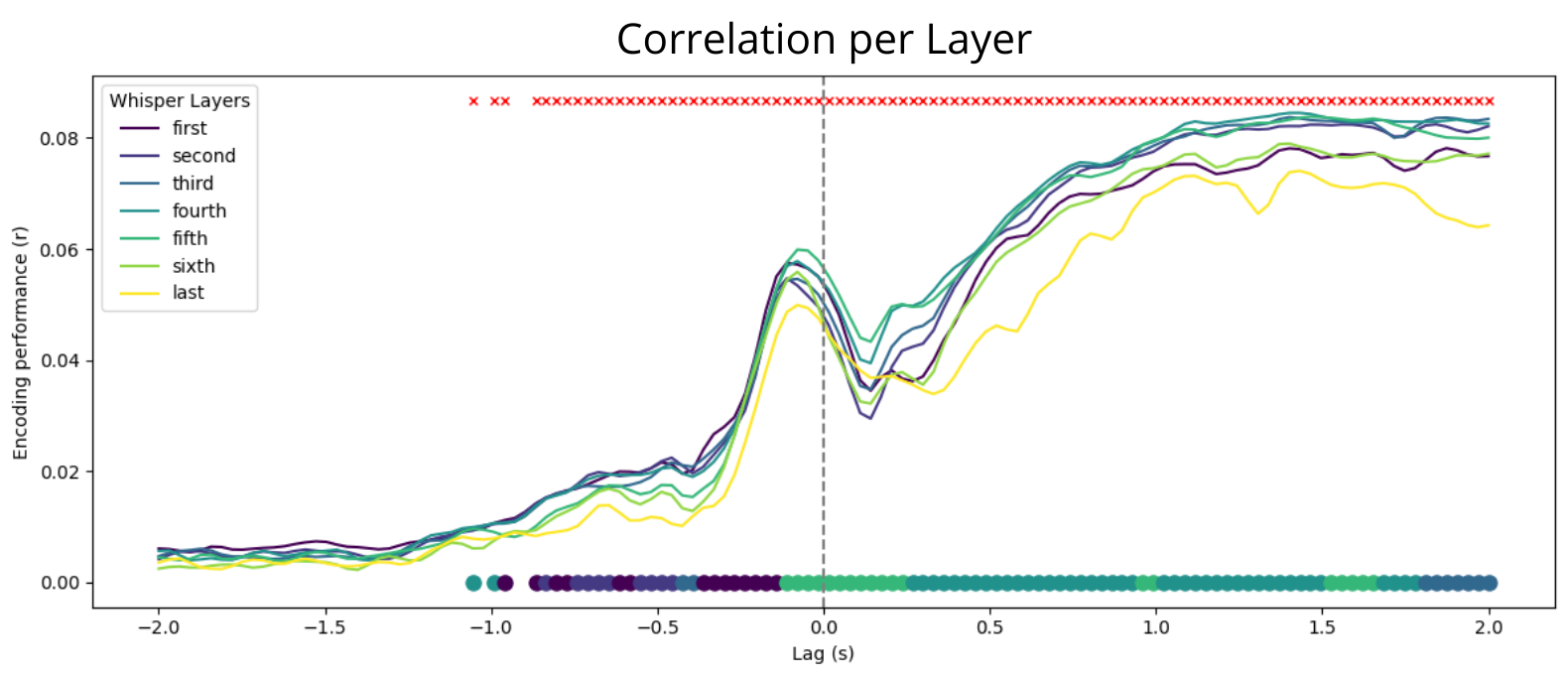}
    \caption{Time-resolved encoding performance across Whisper layers. The curves show the mean Pearson correlation (averaged across channels and subjects) between predicted and observed neural activity. Colored dots along the bottom indicate, for each timepoint, the layer achieving the highest performance. Red crosses along the top mark timepoints that reached statistical significance under a permutation test against a null distribution obtained by shuffling word–neural pairings. Intermediate layers consistently outperform both lower and higher layers.}
    \label{fig:corr_time_layer}
\end{figure*}

Hierarchical representations from Whisper reliably predicted ECoG responses during naturalistic speech perception (Fig.~\ref{fig:corr_time_layer}). All layers exhibited a rise in encoding performance shortly before word onset with a sharp peak followed by a brief dip and a subsequent increase during the post-onset interval. While layer depth is often associated with increasing representational abstraction in speech models, we do not directly manipulate or isolate specific linguistic levels. Accordingly, our results are interpreted as evidence for layer-dependent differences in representational alignment with neural activity \citep{goldstein2025unified}.

\begin{figure*}
    \centering
    \includegraphics[width=.99\linewidth]{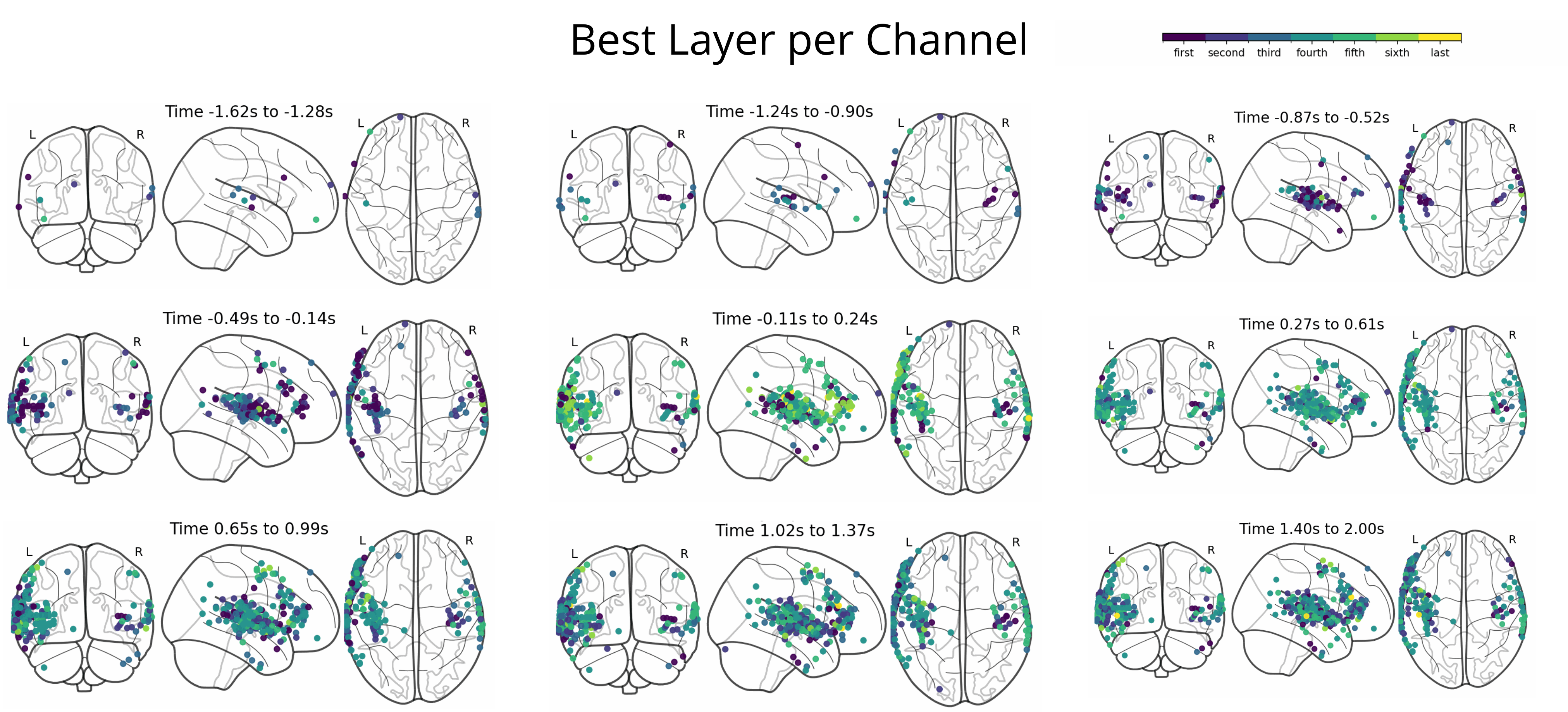}
    \caption{For each electrode, the Whisper layer yielding the highest encoding performance is shown across successive temporal windows relative to word onset. Early windows are dominated by lower-level layers, reflecting acoustic tracking, while later windows show a shift toward middle layers, consistent with phonetic and articulatory processing. Spatial patterns reveal a progression from posterior auditory cortex to more anterior superior temporal and frontal regions, suggesting a temporal-to-linguistic representational gradient.}
    \label{fig:layer_per_channel}
\end{figure*}

Layer-specific analyses showed that intermediate Whisper layers provided the best overall predictions. \textcolor{black}{In Appendix \ref{sec_app:stat_evidence} we show the statistical evidence.} Early layers aligned with pre-onset and onset-related responses, while middle layers dominated after onset, consistent with the timing of acoustic processing (Figure~\ref{fig:layer_per_channel}).

\subsection{Baseline Comparisons}

\textcolor{black}{To contextualize the performance of our proposed encoding model, we compared it against three linear baseline models. First, we used a time-aware linear Whisper baseline based on the same fourth-layer Whisper representations as our model. In this baseline, temporal information was preserved in the input features (flattened across time and feature dimensions), but the mapping from stimulus representations to neural activity was estimated with a linear ridge model, without recurrent dynamics or learned soft alignment. Second, we used a time-averaged Whisper baseline, in which fourth-layer Whisper embeddings were averaged across the temporal window before fitting a linear encoding model. This baseline removes within-window temporal structure and therefore tests how much of the encoding performance can be explained by static word-level acoustic/phonetic representations. Third, we included a text-only GPT-2 baseline, following the approach adopted in the original Podcast ECoG paper~\citep{zada2025podcast}, using middle-layer GPT-2 representations and a linear encoding model.}

\begin{figure*}
    \centering
    \includegraphics[width=.99\linewidth]{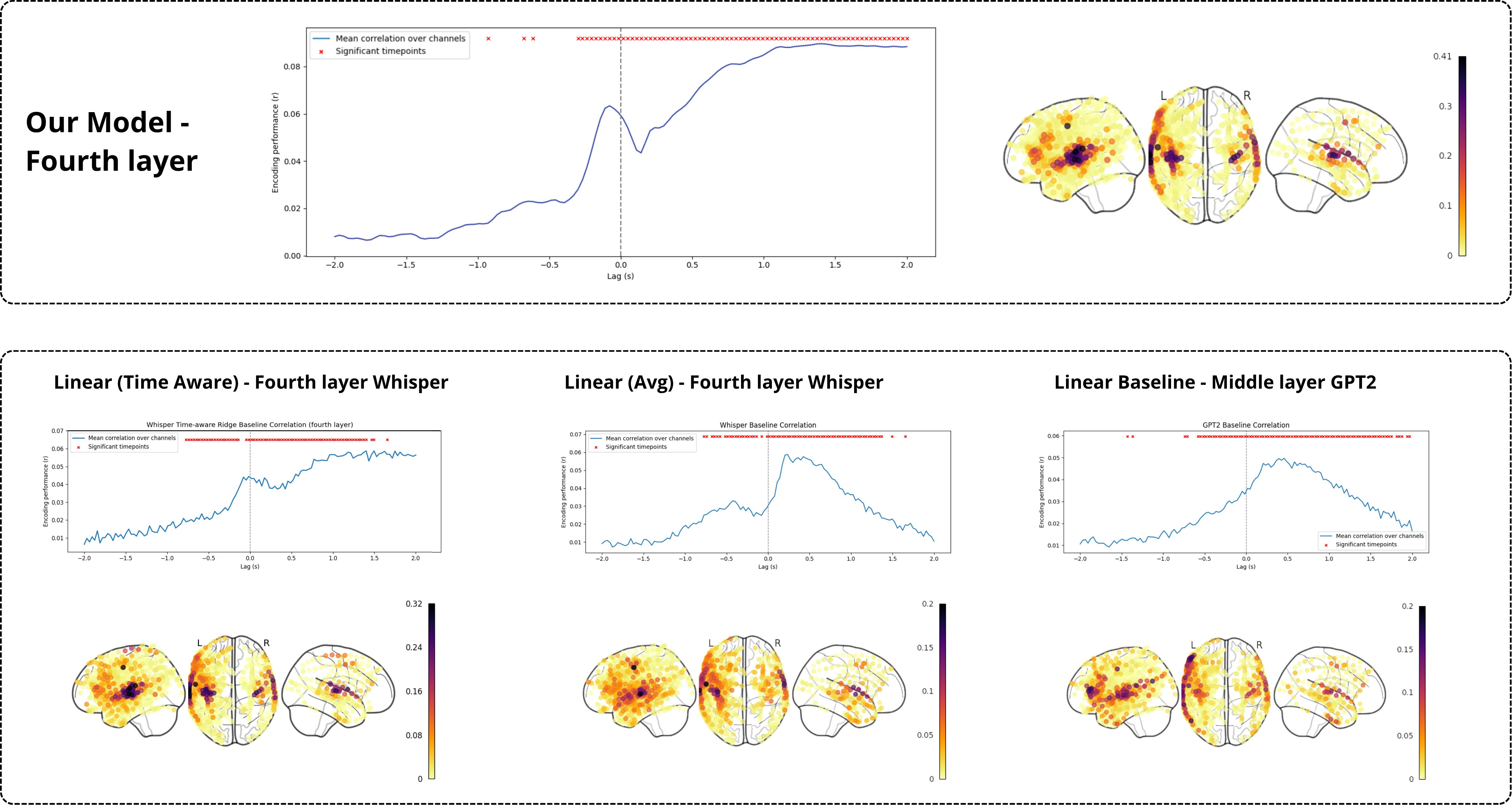}
    \caption{
Comparison between our time-aware encoding model and linear baseline models.
\textbf{Top:} Results obtained with our proposed model using the most predictive Whisper layer (layer~4). The model shows the highest temporal encoding performance and the broadest spatial distribution of predictive electrodes.
\textbf{Bottom left:} Time-aware linear Whisper baseline using the same fourth-layer Whisper representations while preserving temporal information in the input features, but replacing the recurrent/attention-based mapping with linear ridge regression.
\textbf{Bottom middle:} Time-averaged linear Whisper baseline, in which fourth-layer Whisper embeddings are averaged across the temporal window before fitting the linear model.
\textbf{Bottom right:} Text-only GPT-2 linear baseline using middle-layer GPT-2 representations, following the original Podcast ECoG baseline. Across panels, temporal plots show mean encoding performance over selected electrodes, with significant time points marked in red. Spatial maps show electrode-wise encoding performance.
}
    \label{fig:baseline_comparison}
\end{figure*}

\textcolor{black}{Figure~\ref{fig:baseline_comparison} summarizes the temporal and spatial encoding profiles of our model and the three baselines. The proposed model achieved the strongest overall encoding performance and showed the broadest spatial distribution of high-performing electrodes. The time-aware linear Whisper baseline improved over the time-averaged Whisper baseline in magnitude and spatial coverage, indicating that preserving temporal information in the Whisper features is beneficial even for a linear model. The time-averaged Whisper baseline and the GPT-2 baseline both captured significant stimulus-response relationships, but with lower peak correlations and weaker spatial coverage. The time-averaged Whisper baseline preserved a clear post-onset response profile but lacked the richer temporal dynamics observed in the time-aware models.}

The comparison between our model (fourth layer as the highest average result) and the linear Whisper baseline demonstrates that the performance gains are not merely due to the choice of representation, but arise from the explicit modeling of temporal dependencies. By learning a soft-alignment between speech embeddings and neural timepoints, our model captures both anticipatory and post-onset dynamics that are less effectively captured by linear approaches.

\subsection{Interpretability Results}

\begin{figure*}
    \centering
    \includegraphics[width=.99\linewidth]{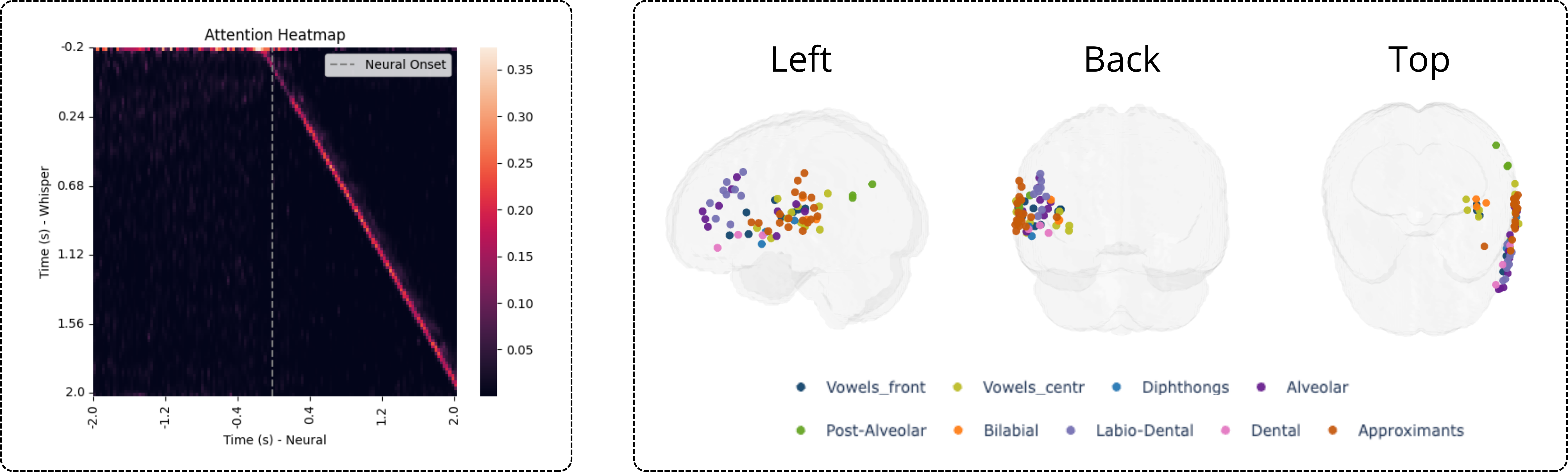}
    \caption{\textbf{Left}: Attention map showing how neural timepoints (horizontal axis, aligned to word onset) weight Whisper encoder representations at different relative temporal offsets (vertical axis). The map is obtained by averaging attention weights across all test samples (i.e., words, electrodes, and subjects). A diagonal structure is observed, indicating a systematic temporal alignment between neural activity and stimulus representations computed over nearby acoustic frames.  \textbf{Right}: Spatial distribution of phoneme-selective electrodes across the temporal lobe. Colors denote broad phoneme categories defined based on articulatory properties (e.g., front/central/back vowels, bilabial, alveolar, dental, approximants).}
    \label{fig:interpretability}
\end{figure*}

The learned attention map in Figure~\ref{fig:interpretability} (left) shows a smooth diagonal structure between Whisper embedding time steps and neural timepoints, indicating that the model tends to place higher weight on input frames that are temporally close to (and often slightly preceding) the neural timepoint being predicted. The observed structure suggests that the model leverages local temporal context in the stimulus representations when mapping speech features to time-resolved neural activity.

The phoneme-selective electrodes identified by our interpretability analysis form anatomically coherent spatial clusters across the temporal cortex (Figure~\ref{fig:interpretability}, right). Although individual phoneme classes partially overlap, their overall organization reveals a systematic structure consistent with known articulatory and acoustic gradients in the human auditory system. Vowel-related categories (front vowels, central vowels, diphthongs) tend to occupy more medial portions of the superior temporal gyrus (STG) and the internal part of the planum temporale. These regions are known to encode spectral and harmonic structure, consistent with the acoustic properties of vowel production \citep{steinschneider2023toward,chang2010categorical}.
Consonant-related categories, by contrast, form denser clusters distributed along the STG, with alveolar, labio-dental, and dental categories appearing more anteriorly, and post-alveolar categories located more posteriorly. Bilabial consonants represent a partial exception, showing a more medial distribution.

Across complementary metrics, the encoding-selected configuration yielded substantially more separable and spatially consistent clusters (Table~\ref{tab:cluster_metrics}): silhouette scores increased markedly, Davies--Bouldin indices decreased, and local neighborhood label entropy was reduced. Bootstrap resampling over electrodes confirmed that these improvements were statistically reliable, indicating that restricting the analysis to encoding-informative channels enhances the anatomical specificity of phoneme-selective organization rather than simply reducing sample size. Consequently, encoding-based electrode selection does not trivially induce spatial clustering, as shown by the markedly lower coherence observed when applying the same pipeline to all electrodes.

\begin{table*}[t]
\centering
\caption{Quantitative comparison of phoneme-cluster spatial coherence using encoding-selected electrodes (most responsive channels) vs.\ all electrodes. Bootstrap statistics are computed by resampling electrodes ($n=1000$) and performing a t-test between distributions.}
\label{tab:cluster_metrics}
\begin{tabular}{lcccc}
\toprule
\textbf{Metric} & \textbf{Selected (obs)} & \textbf{All (obs)} & \textbf{Bootstrap mean$\pm$std} & \textbf{$p$ (one-tailed)} \\
\midrule
Silhouette $\uparrow$ & 0.255 & 0.020 & 0.303$\pm$0.073 vs.\ 0.031$\pm$0.030 & $4.12\times 10^{-8}$ \\
DBI $\downarrow$ & 2.356 & 6.901 & 2.463$\pm$0.586 vs.\ 5.892$\pm$1.477 & $5.67\times 10^{-11}$ \\
Local entropy $\downarrow$ & 1.733 & 1.911 & 1.531$\pm$0.128 vs.\ 1.736$\pm$0.063 & $6.11\times 10^{-5}$ \\
\bottomrule
\end{tabular}
\end{table*}

\section{Discussion}
The primary finding of this work is that intermediate layers of Whisper yield the highest encoding performance when predicting ECoG activity during naturalistic speech perception. This result is consistent with a robust pattern observed across brain–model alignment studies: representations from middle layers of large language and speech models tend to best match neural responses, whereas early layers are dominated by low-level sensory features and deeper layers become increasingly semantic-specific. Similar effects have been reported for text-based foundation models such as GPT-2 and BERT, where intermediate layers most strongly predict fMRI, MEG, and ECoG responses during reading or listening \citep{jain2018incorporating, schrimpf2021neural, caucheteux_brains_2022, goldstein_shared_2022}. Our results extend this principle to a large-scale speech foundation model and to intracranial, time-resolved recordings, supporting the idea that intermediate representational stages capture the level of abstraction most relevant for cortical speech processing.


The learned attention maps provide a useful visualization of how the model aligns Whisper embeddings with neural timepoints. Their diagonal structure indicates that predictions are dominated by temporally local input frames, often slightly preceding the predicted neural response. This pattern is consistent with known auditory cortical latencies and with predictive mechanisms during speech perception, whereby upcoming linguistic information is partially anticipated based on context. However, attention weights should not be interpreted as direct physiological latency estimates. More generally, the relationship between attention and explanation remains debated: attention can provide useful diagnostic information about model behavior, but it is not automatically equivalent to a faithful mechanistic explanation \citep{jain2019attention, lamarre2022attention}.

Beyond encoding accuracy, our phonemic interpretability analysis identifies electrodes selectively responsive to articulatory phoneme categories, forming anatomically coherent clusters across the temporal lobe. This organization is strongly consistent with prior intracranial and electrophysiological evidence showing that STG encodes phonetic and articulatory features of speech. Classic ECoG studies have demonstrated systematic representation of phoneme categories and phonetic features in STG during continuous speech perception \citep{chang2010categorical, mesgarani2014phonetic}, as well as separable neural responses to vowels and consonants. Our results do not introduce new phoneme categories per se, but demonstrate that such cortical structure can be recovered through an encoding model grounded in Whisper representations, linking phoneme-level neural selectivity to representations learned by a modern speech foundation model. 

\subsection{Baseline comparisons}
A question motivating this work was whether the temporal resolution and spatial specificity of intracranial recordings justify the use of a more expressive encoding model than standard linear baselines. This is not a trivial assumption: linear encoding models are often preferred in neural data analysis because they are stable, interpretable, and less prone to overfitting, especially when the number of subjects is limited. However, ECoG high-gamma activity provides a temporally precise and spatially localized measure of cortical responses, making it plausible that the relationship between speech representations and neural activity involves temporally structured and nonlinear dependencies that are difficult to capture with a single linear mapping. The baseline comparisons support this view. The time-averaged Whisper baseline tests whether static word-level information from the same speech representation is sufficient to explain the neural response, whereas the time-aware linear Whisper baseline preserves within-window temporal structure but removes recurrent dynamics and learned soft alignment. The model appears to benefit from learning a flexible, nonlinear mapping between speech-model states and neural timepoints. In this sense, the intracranial setting provides a stronger motivation for temporal model complexity: the additional expressivity is not introduced only for predictive performance, but to test whether high-resolution cortical responses contain structured relationships that linear models capture only partially.

\subsection{Limitations}
Several limitations should be noted. First, the electrode coverage is clinically determined rather than experimentally controlled. As a result, anatomical coverage varies across participants and some regions may be under-sampled. Second, the word-locked windows overlap in continuous time, which introduces temporal dependence between samples; although we address this issue through contiguous cross-validation and subject-level inference, residual autocorrelation may still affect summary estimates. Third, our interpretability analyses are correlational and model-based. Attention maps, layer comparisons, and phoneme-selective clusters provide evidence of structured alignment, but they do not establish causal mechanisms of speech perception. Future work should test whether the same layer-wise and temporal-alignment patterns generalize across larger speech models, other self-supervised encoders, additional datasets, and experimental conditions with controlled acoustic and linguistic manipulations.

\section{Conclusion}
Together, these findings support a convergent picture in which intermediate representations of large speech models provide the best match to cortical speech processing, both in terms of encoding performance and interpretability. The results strengthen the view that foundation models trained on large-scale speech data capture representational stages that are not only useful for recognition, but also aligned with the neural computations underlying human speech perception.


\bibliographystyle{plainnat}
\bibliography{main}

\appendix
\setcounter{table}{0}
\setcounter{figure}{0}
\renewcommand{\thetable}{A\arabic{table}}
\renewcommand{\thefigure}{A\arabic{figure}}
\clearpage

\section{Statements}

\section*{Ethics Statement}
This study uses only publicly available, anonymized intracranial electrophysiology data from the Podcast ECoG dataset (Zada et al., 2025). No new human-subject data were collected. The original dataset was acquired with informed consent and institutional ethical approval; electrode placement was determined solely by clinical needs. References: \cite{zada2025podcast}, Podcast ECoG Dataset, Scientific Data, \url{https://doi.org/10.1038/s41597-025-05462-2}; OpenNeuro ds005574, \url{https://doi.org/10.18112/openneuro.ds005574.v1.0.2};

\section{Supplementary Material}

\subsection{Projection to Glasser-Atlas}

\begin{figure}[h]
    \centering
    \includegraphics[width=\linewidth]{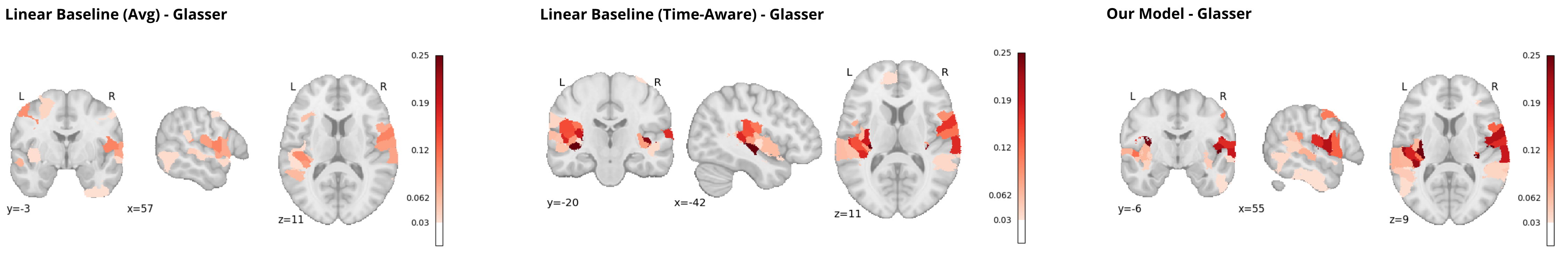}
    \caption{
    ROI-level projection of encoding correlations onto the Glasser atlas.
    \textbf{Left:} linear baseline, Whisper embeddings with time-averaging and linear regression. 
    \textbf{Center:} linear baseline, Whisper embedding flattening timepoint and feature dimensions.
    \textbf{Right:} our time-aware model. Warm colors indicate higher mean Pearson correlation within each ROI (aggregated over electrodes assigned to that parcel; colorbar shared across panels).
    All approaches highlight auditory and peri-auditory cortex, but our model yields stronger and more spatially concentrated ROI responses, with prominent effects in core/belt auditory parcels and adjacent opercular/insular regions.
    }
    \label{fig:glasser_roi}
\end{figure}

To assess whether the encoding effects are anatomically coherent at the ROI level, we projected channel-wise encoding correlations onto the Glasser cortical atlas \citep{glasser2016multi}. Specifically, for each ROI we aggregated the correlations of electrodes falling within that parcel and reported the mean correlation and number of contributing channels (Figure~\ref{fig:glasser_roi}).

\begin{table}[h]
\centering
\scriptsize
\setlength{\tabcolsep}{4pt}
\renewcommand{\arraystretch}{1.05}
\setlength{\belowcaptionskip}{6pt}
\caption{Top Glasser ROIs by mean encoding correlation. For each ROI we report the model, hemisphere (hemi), mean Pearson correlation across electrodes within the ROI, and the number of contributing channels.}
\label{tab:glasser_toproi}
\begin{tabular}{lllc c}
\toprule
\textbf{Model} & \textbf{ROI} & \textbf{Hemi} & \textbf{Mean $r$} & \textbf{\#Ch} \\
\midrule
\multirow{6}{*}{Linear Avg.}
 & Area 43 & L & 0.1439 & 2 \\
 & Primary Auditory Cortex & R & 0.1092 & 4 \\
 & Area 55b & R & 0.1063 & 1 \\
 & Lateral Belt Complex & R & 0.0956 & 5 \\
 & Area OP4/PV & L & 0.0836 & 2 \\
 & Auditory 4 Complex & L & 0.0795 & 5 \\
\midrule
\multirow{6}{*}{Linear Time-Aware}
 & Area 52 & R & 0.2536 & 1 \\
 & Medial Belt Complex & L & 0.2325 & 4 \\
 & Insular Granular Complex & R & 0.1873 & 2 \\
 & Lateral Belt Complex & R & 0.1746 & 5 \\
 & Auditory 4 Complex & L & 0.1712 & 5 \\
 & Area 43 & L & 0.1635 & 2 \\
\midrule
\multirow{6}{*}{Our Model}
 & Area 52 & R & 0.3160 & 1 \\
 & Medial Belt Complex & L & 0.3116 & 4 \\
 & Insular Granular Complex & R & 0.2438 & 2 \\
 & Lateral Belt Complex & R & 0.2237 & 5 \\
 & Area OP1/SII & R & 0.2101 & 2 \\
 & Area 43 & L & 0.2044 & 2 \\
\bottomrule
\end{tabular}
\end{table}

Both the linear Whisper baseline and our model highlight auditory and peri-auditory regions, indicating that the strongest encoding effects are not spatially random. However, our model produces a markedly stronger and more focal ROI pattern: peak correlations are higher and concentrate in core and belt auditory areas (e.g., Area~52, Medial Belt Complex, Lateral Belt Complex), as well as adjacent opercular/insular regions (e.g., OP1/SII, Insular Granular Complex). In contrast, the linear baseline yields lower ROI means and a weaker overall anatomical contrast (Table~\ref{tab:glasser_toproi}), consistent with the reduced time-locked encoding performance observed in the baseline comparison.

\subsection{Intermediate Layers as Statistical Evidence}
\label{sec_app:stat_evidence}

\begin{figure*}[h]
    \centering
    \includegraphics[width=.90\linewidth]{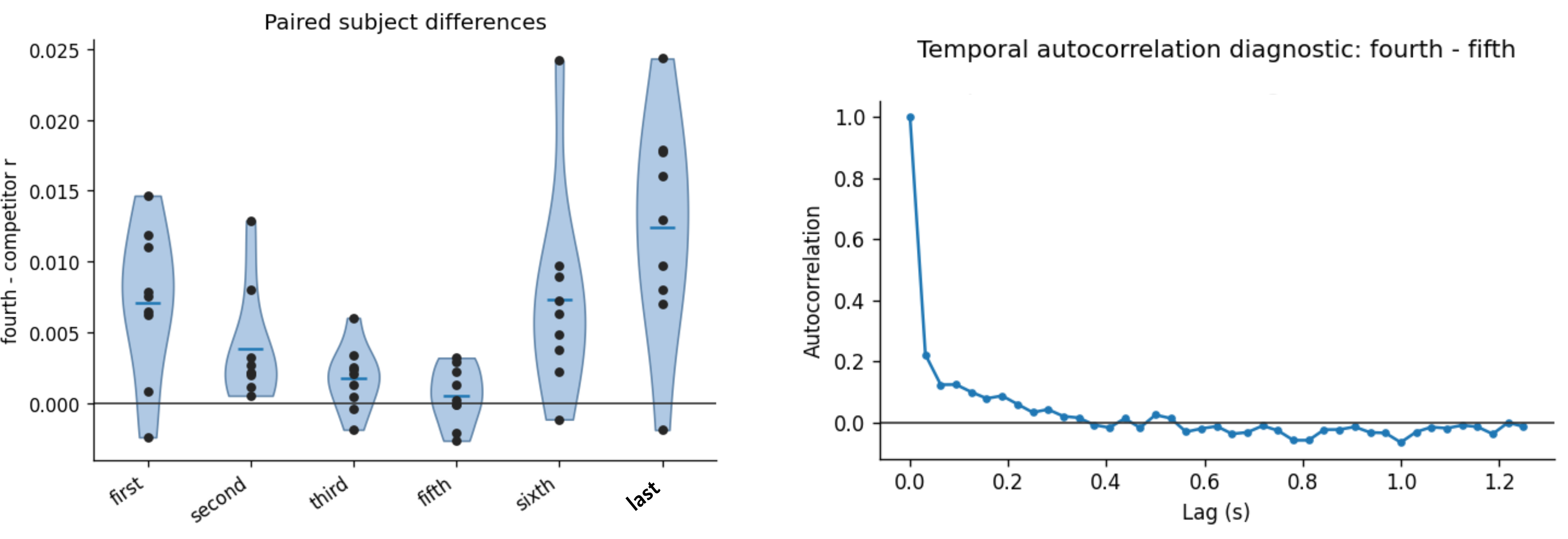}
    \caption{Subject-level layer comparison and temporal-dependence diagnostic. Left: paired differences in encoding performance between Whisper layer 4 and each competing layer. Each dot represents one subject. Right: temporal autocorrelation of the subject-level layer 4 minus layer 5 difference curve, showing short-range autocorrelation between adjacent time points.
    }
    \label{fig:statistical_evidence}
\end{figure*}

\textcolor{black}{To make the layer-comparison analysis more conservative, we repeated the inference using subjects as the statistical unit. For each subject and each Whisper layer, encoding performance was first averaged across cross-validation folds, electrodes, and time points within the pre-specified analysis window. This yielded a single summary value per subject and layer. We then computed paired subject-level differences between layer 4 and each competing layer, and tested whether these differences were consistently greater than zero across subjects using paired sign-flip inference, with FDR correction across the planned layer comparisons (Figure \ref{fig:statistical_evidence} left). This analysis showed that layer 4 outperformed layer 1 ($q_{\mathrm{FDR}} = 0.0117$), layer 2 ($q_{\mathrm{FDR}} = 0.0127$), layer 3 ($q_{\mathrm{FDR}} = 0.0304$), layer 6 ($q_{\mathrm{FDR}} = 0.0119$), and the final layer ($q_{\mathrm{FDR}} = 0.0110$). In contrast, the difference between layer 4 and the adjacent layer 5 was small and not significant ($q_{\mathrm{FDR}} = 0.2183$).}

\textcolor{black}{Importantly, this averaging procedure does not eliminate temporal autocorrelation within the original time-resolved encoding curves. Adjacent time points remain correlated, as expected given the use of overlapping temporal windows. We therefore explicitly quantified this dependence by computing the temporal autocorrelation of the subject-level difference curve for the closest competing contrast, layer 4 minus layer 5 (Figure \ref{fig:statistical_evidence} right). However, the revised inference removes the corresponding sample-independence problem by not using time points as statistical replicates. Temporal samples are summarized within each subject before hypothesis testing, so the final statistical test is performed over subjects rather than over time points. Thus, temporal autocorrelation may affect each subject's summary estimate, but it cannot inflate the nominal sample size or the degrees of freedom of the layer-comparison test.}

\subsection{Temporal Ablation}

In this chapter we address the presence of encoding performance at timepoints distant from word onset, potentially reflecting temporal leakage induced by the bidirectional architecture or by the choice of temporal context. We conducted two complementary ablation analyses targeting (i) architectural temporal information and (ii) input temporal windowing.

\begin{figure*}[h]
    \centering
    \includegraphics[width=.80\linewidth]{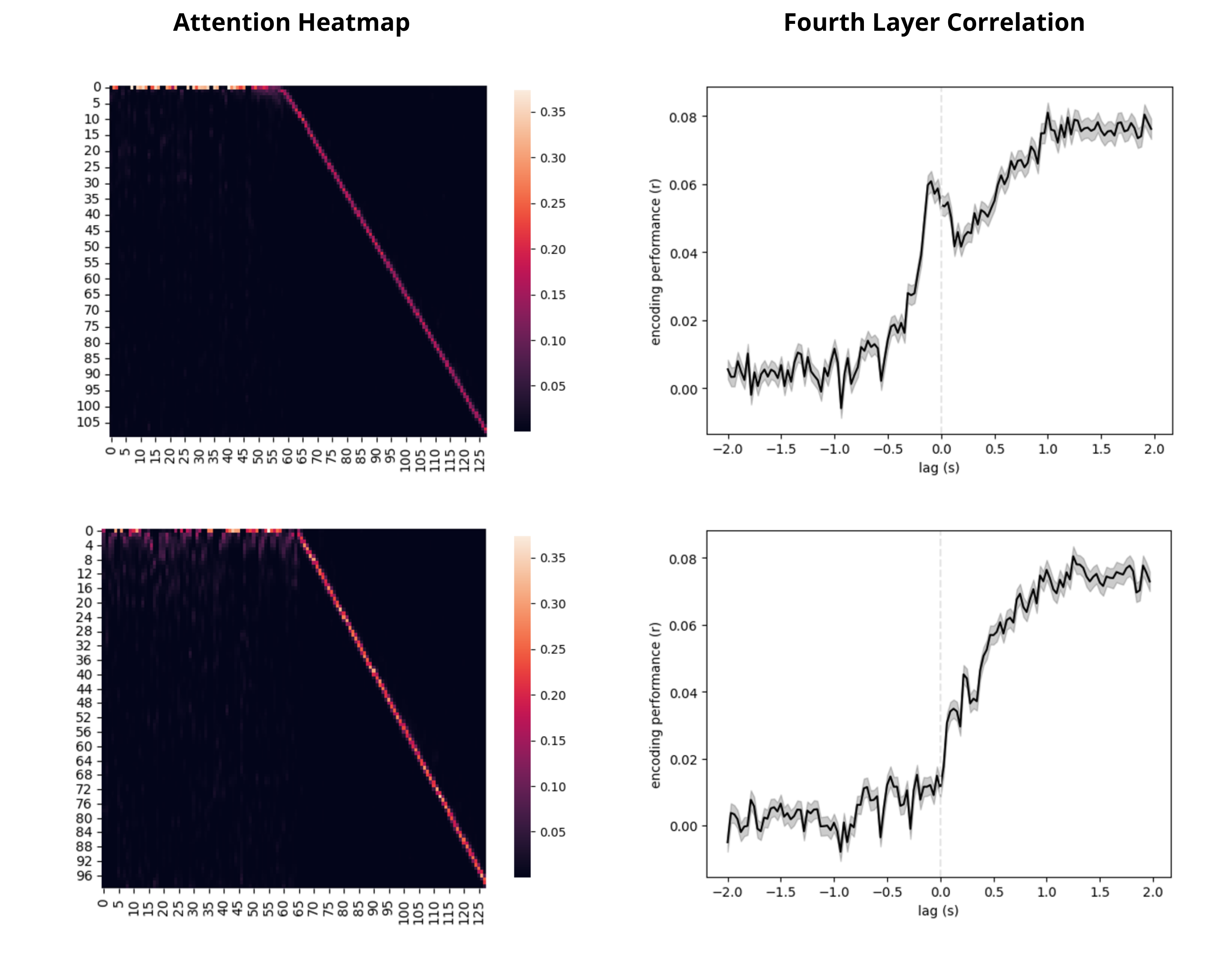}
    \caption{
Temporal ablation analysis. Left: average temporal attention maps showing the alignment between input speech representations and predicted neural timepoints. Right: encoding performance. Top: unidirectional GRU model with pre-onset context (200 ms). Bottom: same GRU model without pre-onset context. The removal of pre-onset input eliminates the pre-onset peak in encoding performance, while preserving the post-onset dynamics.
}
    \label{fig:tempo_ablation}
\end{figure*}

To assess whether future information contributes to the observed encoding patterns, we replaced the bidirectional GRU with a strictly causal (unidirectional) GRU, preventing the model from accessing future input frames when predicting neural responses. Results show that the overall temporal profile of encoding performance remains qualitatively unchanged (Figure~\ref{fig:tempo_ablation} first row). In particular, the post-onset rise in correlation and the sustained encoding performance are preserved under the causal architecture. This indicates that the model does not rely on future information to achieve its performance, and suggests that the observed temporal structure reflects genuine alignment between stimulus representations and neural activity.

A second potential source of anticipatory effects is the inclusion of pre-onset acoustic context (200 ms before word onset) in the input. To evaluate its impact, we repeated the analysis removing any pre-onset audio, such that the model only receives information starting from word onset. 

As shown in Figure~\ref{fig:tempo_ablation} line (b), removing pre-onset context eliminates the peak in encoding performance observed before word onset. The resulting temporal profile becomes strictly post-onset, with encoding performance rising only after stimulus presentation. This result demonstrates that the pre-onset peak observed in the main model is not an artifact, but rather reflects the availability of preceding acoustic context. Importantly, this context corresponds to natural speech continuity, where upcoming words are partially predictable from prior linguistic input.

\subsection{Hyperparameter selection}
\label{app:hyperparameter_selection}

\textcolor{black}{Table \ref{tab:hyperparameters} reports the hyperparameters used for the neural encoding model and for the phonemic clustering analysis. We focus on parameters that directly affect the temporal alignment, model capacity, optimization procedure, activation selection, and spatial refinement of phoneme-selective electrodes. }

\begin{table}
\centering
\small
\caption{Main hyperparameters used for the neural encoding model and the phonemic clustering analysis.}
\label{tab:hyperparameters}
\begin{tabular}{lll}
\hline
\textbf{Block} & \textbf{Hyperparameter} & \textbf{Value (best)} \\
\hline
\multicolumn{3}{l}{\textit{Neural encoding model}} \\
\hline
Feature extractor & Whisper model & \texttt{openai/whisper-base}, frozen \\
Input window & Audio window & $-0.2$ s to $+2.0$ s \\
Target window & Neural response window & $-2.0$ s to $+2.0$ s \\
Architecture & GRU hidden size & 256 \\
Architecture & Num. Layers & 1 \\
Architecture & Dropout & 0.0 \\
Training & Batch size & 64 \\
Training & Learning rate & $10^{-4}$ \\
Training & Weight decay & $10^{-3}$ \\
Loss & Loss function & Temperature-scaled contrastive cosine loss \\
Loss & Learnable temperature init. & 0.03 \\
Cross-validation & Split strategy & Contiguous KFold, 4 folds \\
\hline
\multicolumn{3}{l}{\textit{Phonemic clustering analysis}} \\
\hline
Activation selection & Data used & Observed ECoG responses; encoding-selected channels \\
Activation selection & Z-scoring & Per channel \\
Activation selection & Z threshold & 2.3 \\
Phoneme conversion & G2P model & \texttt{g2p\_en} \\
Phoneme filtering & Removed tokens & \texttt{SIL}, \texttt{SPN}, blanks \\
Category assignment & Statistical test & G-test vs. global phoneme-category distribution \\
Multiple comparisons & Correction / alpha & Benjamini--Hochberg FDR / 0.05 \\
Spatial refinement & Neighbor radius & 20 mm \\
Spatial refinement & Label rule & Majority vote among significant neighboring labels \\
Cluster evaluation & Metrics & Silhouette, Davies--Bouldin, local label entropy \\
\hline
\end{tabular}
\end{table}

\end{document}